# ROTATIONAL DRAG FORCE: A VIABLE ALTERNATIVE TO DARK MATTER

Max I. Fomitchev

## ABSTRACT

I present an alternative explanation of flat rotational curves of galaxies that does not require dark matter but rather relies on classical Newtonian dynamics and an overlooked effect of quantum tunneling. I introduce a rotational drag force, which arises from subatomic particle tunneling through potential barriers caused by energy fluctuations in rotating gravitational field, and present a model for a galactic rotational curve based on Newtonian dynamics, rotational drag force and galactic mass-density distribution.

To support the rotational drag force hypothesis I supply the results of analytical modeling and numerical simulation of effects of the rotational drag force on stellar orbits and galactic morphology. The obtained results provide a clear indication that the rotational drag force can be successfully applied to explaining a wide variety of observational phenomena ranging from flat rotational curves of galaxies and Tully-Fisher relation to origination of spiral galactic arms, galactic bars, and galactic warps as well as anomalous Pioneer 10/11 acceleration and possibly high-velocity galactic clouds.

*Subject headings*: cosmology: dark matter – galaxies: spiral –
galaxies: kinematics and dynamics

## 1. INTRODUCTION

Directly applied Newtonian gravitational dynamics fails to yield correct predictions for stellar orbital velocities in galaxies. Observed rotational velocities of stars in galaxies typically exceed those calculated by means of Newtonian dynamics for the observed luminous galactic mass with excessive velocities, which are manifested by flat rotational curves, being commonly attributed to the existence of dark matter.

Yet today there is little reason to doubt Newtonian dynamics, which has been very successful and accurate in its predictions on the scale of Solar system. Therefore to preserve the Newtonian dynamics on galactic scale the existence of non-luminous and hence not directly observed matter, which contributes to Newtonian gravitational potential and gives rise to stellar velocities, is assumed. There are, however, profound problems with all dark matter theories. Most notably, no convincing rotational velocity fall-off has been observed, which presumes rather peculiar distribution of dark matter and abnormal disproportion of dark matter in comparison to luminous matter, and no successful dark matter candidate has been found. Massive compact halo object (MACHO) counts derived from the observed microlensing events account only for a small fraction of dark matter halo of Milky Way (Rahvar 2003), while weakly-interacting massive particles (WIMPs) such as Axions, massive neutrinos or a variety of super-symmetric particles that are thought to comprise the lion share of the dark halo are yet to be convincingly detected (for a current overview of WIMP detection effort see e.g. Morales 2002).

Perhaps this elusive nature of dark matter has led Milgrom to believe that Newtonian dynamics should be modified to yield a form compatible with observed flat rotational curves, which implies acceleration dependence on distance of $a(r) = GM/r$, rather $a(r) = GM/r^2$ as prescribed by Newtonian dynamics (Milgrom 1983). To reconcile the modified dynamics (MOND) with results successfully predicted by classic Newtonian dynamics (e.g. stellar dynamics on the scale of Solar system) Milgrom postulates that the modification occurs only for small accelerations, $a < a_0$. Perhaps the artificial nature of Milgrom's acceleration-limiting construct coupled with the lack of underlying basis or convincing reasoning behind MOND (except for a drive to explain the observational data) and some published criticism (e.g. Mortlock & Turner 2001) has kept MOND and its followers in minority for 20 years.

## 2. ROTATIONAL DRAG FORCE

### 2.1. Refutation of Ideal Sphere Approximation

In the realm of Newtonian mechanics stellar bodies are typically treated as point masses or ideal spheres. I submit that ideal sphere approximation is insufficient for rotating bodies because the rotation of an ideal sphere results in a logical impossibility.

An ideal sphere would have neither surface defects nor internal structure that would allow us detecting its rotation either visually of by conducting any sort of experiment. Indeed, a rotating ideal sphere would not produce any surface brightness variation or surface friction (because it is ideally smooth and uniform), nor it will produce a magnetic field for a current is required to generate such field, and introduction of such current would violate the idealness of the sphere. Lastly, it is impossible to detect angular momentum of a rotating ideal sphere because it is impossible to turn an ideal sphere around due to the absence of surface friction. Therefore I conclude that an ideal sphere (or point mass) approximation is fundamentally flawed when rotating bodies are concerned as we overlook the very reason that makes rotation detectable - non-uniformities in the structure of the rotating body, which as I will show below have profound effect on the resulting gravitational field.

### 2.2. Non-Uniform Rotating Sphere

I propose a *non-uniform* sphere as a more accurate approximation of a rotating body, where the term *non-uniform* means detectable internal structure such as atomic or particle composition. Particle composition implies that a body is comprised out of a large number of microscopic ideal spheres (e.g. atoms or other particles). The individual properties of such microscopic ideal spheres are not important except for their ability to induce gravitational field of a point mass.

When gravitational field of a rotating spherical stellar body is modeled as a sum of gravitational fields of point masses comprising the body, the resulting gravitational field microstructure exhibits microscopic grooves corresponding to variations in gravitational field energy, fig. 1. The depth of each groove is proportional to the mass of the body (i.e. the number of individual point masses or atoms contributing to each groove) while the width of each groove is proportional to the mass-density of the body.

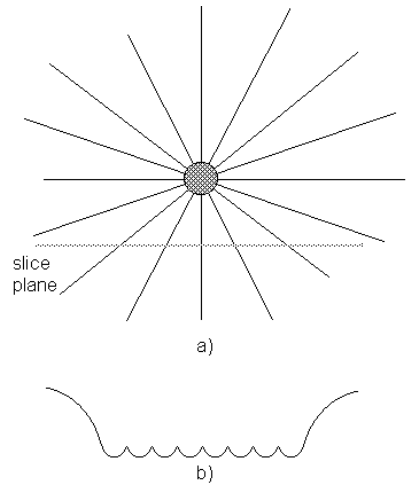

Fig. 1.-Conceptual sketch of groove-like gravitational field energy variations produced by non-uniform gravitating sphere: a) view from the top; b) 'slice' through the gravitational field.

Taking into account the fact that the gravitational field produced by non-uniform sphere in real world is superimposed on top of the background gravitational field of the Universe, which is a subject quasi-random fluctuation, the groove-like gravitational field structure becomes a mesh-like gravitational field structure, which rotates with (or being dragged by) the central body.

### 2.3. Tunneling

When a non-uniform gravitating body rotates, the mesh-like gravitational energy fluctuations produced by its gravitational field sweep through space, rotating around the body's center of mass. The effect of sweeping energy fluctuations on other objects in vicinity of the rotating body cannot be neglected. Due to the extreme minute level of such fluctuations their effect should be considered on microscopic level. Since material bodies in the vicinity of a rotating sphere are also comprised of atoms or particles, the process of sweeping gravitational field fluctuations on microlevel is equivalent to a collision of ensembles of potential barriers arising from energy fluctuations in the gravitational field with ensembles of particles comprising a body. Thus, given the proven rules of quantum mechanics, it is safe and reasonable to expect that individual particles would either tunnel through the approaching potential barriers or reflect from them, depending on particles' momentary kinetic energy.

Concluding from reasoning given in the preceding paragraph, the statistical effect of gravitational energy fluctuations sweeping through a body is manifested by a *drag force* being excerpted on the body. For a macroscopic body the drag force is proportional to the particle reflection probability from potential barriers caused by gravitational energy fluctuations:

$$a_1 = a_0 R \qquad (1)$$

where $a_1$ – is the acceleration of the body being dragged, $a_0$ – acceleration of the dragging body (i.e. central mass), and $R$ – is the reflection probability from the potential barrier, which is statistically averaged for the macroscopic body.

In the extreme case when the gravitational fluctuations are large enough and the comprising particles' kinetic energy is correspondingly low, 100% reflection from the approaching potential barriers will occur, and the body will be dragged in unison with sweeping gravitational field fluctuations. On another extreme, when the gravitational energy fluctuations are very small or the particle energy is large, 100% tunneling will occur and the drag force will be non-existent.

### 2.3. Rotational Drag Force Formula

Given a rotating stellar body, which corresponds to a non-uniform sphere, we are now dealing with two forces, classical Newtonian force and an additional rotational drag force, which has not been accounted for previously:

$$a = a_N + a_{rot} \qquad (2)$$

To simplify the forthcoming discussion I will focus on circular orbits described by radius-vector $r$, residing in plane of the central body rotation. Thus I obtain the following formula for rotational drag force acceleration:

$$a_{rot}(r) = \overline{w}^2 rR = \frac{(v - wr)^2}{r}\big(1 - D\big) \qquad (3)$$

where $\overline{w}$ – is the angular velocity of the difference between the rotation of the central body and the orbiting body; $v$ – orbital velocity; $D$ – potential barrier tunneling probability ($D = 1 - R$).

The expression for the rotational drag force acceleration implies that no drag force is exerted on the orbiting body when it co-rotates with the central body, while angular velocities above or below co-rotation will result in imminent drag force action.

Perhaps the most interesting component of the drag force acceleration is tunneling probability $D$, which is expressed in classical quantum-mechanical terms as

$$D = 1 - R = 1 -$$

$$\left[1 + \frac{U_0^2}{4E(U_0 - E)}\sinh^2\left(\frac{l}{\hbar}\sqrt{2m(U_0 - E)}\right)\right]^{-1} \qquad (4)$$

where $U_0$ – is the potential barrier height, E – is the particle kinetic energy, $l$ – barrier width and $m$ – particle mass.

Assuming non-relativistic velocities, statistically averaged particle kinetic energy of circular orbital rotation measured relatively to the approaching potential barrier is given by

$$E = \frac{m\bar{v}^2}{2} = \frac{m(v - \boldsymbol{w}r)^2}{2} \tag{5}$$

Now, considering that

$$\sinh(x) \approx x, \; x \ll 1 \tag{6}$$

we can simplify the expression for $D$ the following way:

$$\frac{U_0^2}{4E(U_0 - E)} \sinh^2\left(\frac{l}{\hbar}\sqrt{2m(U_0 - E)}\right) \approx$$

$$\frac{U_0^2}{4E(U_0 - E)} \frac{l^2}{\hbar^2} 2m(U_0 - E) =$$

$$\frac{l^2}{\hbar^2} \frac{U_0^2 m}{2E} = \frac{l^2}{\hbar^2} \frac{U_0^2}{\bar{v}^2} = \frac{1}{\boldsymbol{g}^2 \bar{v}^2} \tag{7}$$

where

$$\boldsymbol{g}^2 \equiv \frac{\hbar^2}{l^2 U_0^2} \tag{8}$$

Thus

$$1 - D = 1 - \frac{1}{1 + \dfrac{1}{\boldsymbol{g}^2 (v - \boldsymbol{w}r)^2}} = \frac{1}{1 + \boldsymbol{g}^2 (v - \boldsymbol{w}r)^2} \tag{9}$$

and we arrive at final formula for rotational drag force acceleration:

$$a_{rot}(r) = \frac{(v - \boldsymbol{w}r)^2}{r(1 + \boldsymbol{g}^2 (v - \boldsymbol{w}r)^2)} \tag{10}$$

## 2.4. Gravitational Field Potential Barrier Energy, $U_0$

The validity of expression (10) rests on the assumption (6) that

$$\frac{l}{\hbar}\sqrt{2m(U_0 - E)} \ll 1 \tag{11}$$

The expression above holds ground presuming that the rotational drag force couples to quarks confined in normal baryonic matter (e.g. protons and neutrons).

Indeed, supposing that the rotational drag force couples to matter on quark level one can expect potential barrier width $l$ to be on the order of proton radius or ~$10^{-15}$ m or smaller.

Particle mass $m$ should be on the order of quark mass level, ~1MeV or $10^{-30}$ kg (u-quark mass is thought to be ~5 MeV).

Particle kinetic energy $E$ can not exceed proton rest mass-energy, $m_p$ ~1GeV or $10^{-10}$ J and certainly must be much smaller than $m_p$ since there are three quarks locked in proton and much of the proton mass-energy is though to come from the strong interaction 'confinement' energy. Therefore I consider quark rest mass-energy, ~1MeV = $10^{-13}$ J to be a good estimate of $E$.

Lastly, the gravitational field potential barrier energy $U_0$, must be on the order of $E$ to assure non-zero tunneling probabilities. In other words if $U_0 \gg E$ tunneling will never occur and the drag force will be zero; if $U_0 \gg E$ the tunneling probability will reach 100% making the rotational drag force much stronger than gravity, which is clearly not the case. Therefore the only reasonable expectation for $U_0$ is $U_0 \sim E$.

Substituting for the proposed values in expression (11) we obtain:

$$\frac{10^{-15}}{6.6 \cdot 10^{-34}}\sqrt{10^{-30}10^{-13}} \approx 10^{-2} \ll 1$$

The assumption (11) holds.

I would like to note that there's got to be a variation of $U_0$ with distance. It is reasonable to expect that the gravitational field fluctuations diminish in magnitude with the increase in distance $r$ from the gravitating body: similarly to blowing an uneven rubber bubble the bumps will get smaller as the bubble radius grows. The same logic, however, dictates that the potential barrier width $l$ will increase with distance similarly to the linear dimensions of the bumps on bubble's surface. Thus the increase in $l$ is likely to cancel out the decrease in $U_0$ at least to a certain degree.

Therefore I conclude that no rapid decline of tunneling probability with distance is expected.

### 2.5. Non-universality of Rotational Drag Force

Tunneling probability $D$ is largely determined by $U_0$, the only 'free' parameter in expression (4). In the absence of masses $U_0$ either remains constant or represents a very slow declining function of $r$. The introduction of other gravitating bodies will likely affect the value of $U_0$ by the relativistic space-time stretch. In other words gravitational field of external bodies will diminish $U_0$ in some proportion to the 'strength' of the external gravitational field (the value of gravitational potential energy of external field or the magnitude of space-time curvature). One important implication of this assumption on rotational drag force is that bodies that themselves produce substantial gravitational field are subject to reduced rotational drag force in comparison with their low-mass counterparts. I will refer to this phenomenon as *non-universality* of the rotational drag force as it affects low-mass bodies stronger and more massive bodies weaker.

The hypothesis of non-universality of rotational drag force is not without ground. Recent indications of Pioneer 10/11, Ulysses and Galileo space craft anomalous acceleration of ~8.5 x $10^{-10}$ m sec$^{-1}$ directed towards the Sun (Anderson et al. 1998) can be attributed to the rotational drag force induced by the Sun, acting on the spacecraft. Anderson et al. (1998) rules out all conventional explanations for the anomalous acceleration including possible dark matter confined in the vicinity of the Sun noting that 3 x $10^{-4}$ $M_\Theta$ dark matter is necessary to produce the anomalous acceleration, while the accuracy of planetary ephemeris allow only on the order of $10^{-6}$ $M_\Theta$ of dark matter within the orbit of Uranus (Anderson, Lau, & Krisher 1995).

Non-universality of rotational drag force may explain the anomalous acceleration of Pioneer 10/11 spacecraft: the spacecraft have no appreciable mass and therefore produce a minuscule internal gravitational field while the planets are massive enough to reduce the $U_0$ of the Sun to the level of tunneling probability close to a 100%. Therefore the effect of solar rotational drag force is limited to low-mass bodies such as spacecrafts, comets and small asteroids. The latter observation may provide a clue to why primordial comets in solar vicinity are limited to the postulated Kuiper belt / Oort cloud far on the outskirts of the Solar system: they were driven their not just by the extinction due to accretion and gravitational perturbations of gas giants, but also by the rotational drag force of the Sun.

### 2.6. High-Velocity Clouds

Another possible evidence supporting the argument of non-universality of the rotational drag force comes from observation of high-velocity clouds (HVCs) in Milky Way galaxy. HVCs are composed primarily of neutral hydrogen and move about the galaxy with velocities between 80 and 400 km sec$^{-1}$ (Putman 1999), which are clearly incompatible with a simple model of differential galactic rotation (Wakker & van Woerden 1997). Also, low metalicity of HVCs suggests that the clouds are primordial leftovers of the Milky Way and its neighbors (Wakker 1999).

Considering the extremely old age of the clouds it is reasonable to conclude that the HVCs had enough time to accelerate to their present rotational velocities that peak out at values 50-100% higher than rotational velocity of stars in the galactic disk. Unlike stars gas clouds produce less appreciable gravitational field (in terms of space-time curvature) due to their extreme low density. Thus gas clouds are subject to a stronger rotational drag force and therefore they will eventually attain higher rotational velocity than stars in the galactic disk.

### 3. ROTATIONAL VELOCITY MODEL

Now, once we have finished deriving the expression for rotational drag force acceleration, we can derive an expression for rotational velocity $v(r)$ as a function of radius-vector $r$. Once again, assuming circular orbits in plane of the central body rotation total orbital acceleration is given by:

$$a(r) = a_N(r) + a_{rot}(r) \qquad (12)$$

Keeping in mind that

$$a(r) = \frac{v^2}{r} \qquad (13)$$

final rotational velocity formula is

$$v^2 = \frac{(v - \mathbf{w}r)^2}{1 + \mathbf{g}^2(v - \mathbf{w}r)^2} + a_N(r)r \qquad (14)$$

which for Newtonian point-mass approximation translates into

$$v^2 = \frac{(v - \mathbf{w}r)^2}{1 + \mathbf{g}^2(v - \mathbf{w}r)^2} + \frac{GM}{r} \qquad (15)$$

Because he expression (15) is a fourth-degree polynomial the exact analytical solutions for $v(r)$ can be obtained. I have calculated the exact solutions for $v(r)$ using *Wolfram Mathematica* software package finding only two real-value solutions corresponding to the cases when the orbiting body rotates in the same direction as the central body and counter to the central body direction of rotation.

Unfortunately, the exact analytical forms for $v(r)$ obtained using *Mathematica* are several pages long, which make them impossible to comprehend or publish. Instead it is far easier to analyze the expression (15) directly. It is clear that

$$v \to \mathbf{g}^{-1} \text{ with } r \to \infty \qquad (16)$$

The limit $v \to \mathbf{g}^{-1}$ is the case when and $a_N(r) \to 0$ faster than $r^{-1}$, which is a reasonable expectation for realistic central mass-density distributions.

Recalling expression (8) we obtain

$$v \to \frac{lU_0}{\hbar} \qquad (17)$$

The expression (17) imposes another constraint on the value of $U_0$. Considering that for most spiral galaxies rotational curves flatten-out at 200 km sec$^{-1}$, I obtain that product of $lU_0$ must be on the order of $10^{-28}$ J m, which is compatible with the hypothesis that $U_0$ is on the order of $10^{-13}$ J ($l$ is assumed to be on the order of proton radius, $10^{-15}$ m).

$U_0$ is a function of the rotating body mass and density: $U_0 \sim$ mass and $U_0 \sim$ density$^{-1}$. Indeed, more massive bodies would producer 'deeper grooves' in gravitational field and as a result higher potential barriers, whereas denser bodies would produce thinner barriers (smaller $l$). Therefore

$$v_{max} \sim M \qquad (18)$$

where $M$ is the mass of the rotating central body. The relation (18) provides a basis for understanding Tully-Fisher relation. Assuming that the galactic surface brightness is characteristic (and proportional) to the total galactic mass, rotational drag force forces maximum rotational velocity up in proportion to the total galactic mass. Therefore galaxies with greater surface brightness (mass) *must* have greater rotational velocities.

Another observation that we can make my studying the relationship (17) is that extremely dense bodies are expected to exert little or not drag force as $l \to 0$ with increasing density. Therefore neutron stars and most notably black holes should produce drag force lesser than those produced by stars of comparable masses. Even more curiously, the "black holes have no hair" conjecture, which for our purpose implies that a black hole cannot have an internal structure by which it can be characterized, presumes that a black hole cannot produce a drag force at all (no internal structure, finite mass, but infinite mass density). Therefore the "black holes have no hair" conjecture can be used to sort out black hole candidates by analyzing the manifestations of the drag force (if any) induced by a rotating black hole candidate object. On the other hand detection of rotational drag force produced by black hole candidates can disprove the conjecture at least in part, living out the zero electric charge portion of the conjecture, which has no immediate connection to the rotational drag force.

### 4. ROTATIONAL CURVE OF A SPIRAL GALAXY: ANALYTICAL MODEL

Having finished general discussion on the rotational drag force I would like to present the results of analytical rotational curve modeling for a realistic normal spiral galaxy of type (e.g. of type S0, with spiral arms density perturbation negligible). For modeling purposes it is sufficient to consider such spiral galaxy consist of two major components: bulge, modeled as oblate spheroid with Gaussian-like mass-density distribution (see Dwek et al. 1995):

$$\boldsymbol{r}_{bulge}(R) = \frac{\boldsymbol{r}_0}{R}\exp[-\frac{1}{2}R^2] \qquad (19)$$

$$R^2 = \frac{x^2 + y^2}{r_0^2} + \frac{z^2}{z_0^2} \equiv \frac{r^2}{r_0^2} + \frac{z^2}{z_0^2} \qquad (20)$$

and exponential disk derived from the model presented by Wainscoat et al. 1992:

$$\boldsymbol{r}_{disk}(r,z) = \boldsymbol{r}_1\exp[-\frac{r}{h} - \frac{z}{h_z}] \qquad (21)$$

Extensive experimentation with the rotational curve model described by expression (14) using various forms for Newtonian acceleration component $a_N$ reveal that the model is relatively insensitive to the particular analytic form of $\boldsymbol{r}_{disk}$, $\boldsymbol{r}_{bulge}$ and as a result $a_N$, as long as mass-densities fall-off reasonably fast in some Gaussian or exponential-derived form.

The general findings are the following: large massive bulges and low-mass disks tend to create falling-to-flat rotational curves (fig. 2a) while less massive bulges and more massive disks produce rising rotational curves (fig. 2b) with anything in between possible for other ratios of bulge mass-density to the disk mass-density.

Another realistic-looking rotational curve, shown on fig. 3, was modeled for a Milky-Way-size spiral galaxy with both massive bulge and significant disk.

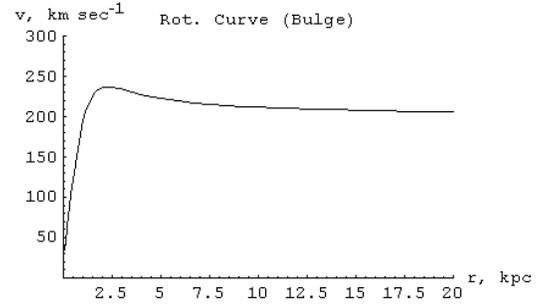

(a)

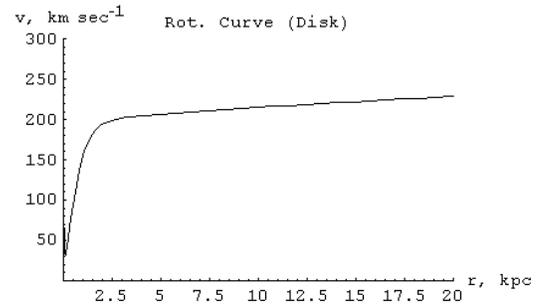

(b)

Fig. 2.-Two common types of rotational curves: a. *falling-to-flat* rotational curve of a galaxy with dominant bulge and b. *rising* rotational curve of a galaxy with dominant disk.

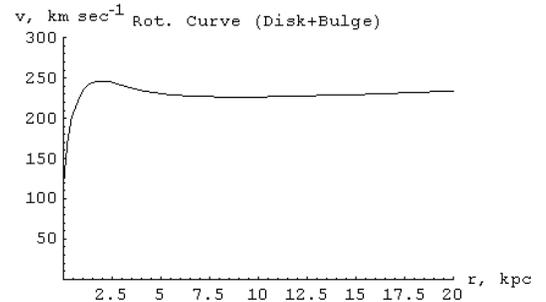

Fig. 3.-Modeled rotational curve of a Milky-Way type spiral galaxy with massive bulge and significant disk..

Thus the rotational drag-force theory allows discerning the dominant component of the galactic structure: rising rotational curves like for (for instance like NGC300, NGC2403, see Kent 1986 and Kent 1987) are characteristic of dominant disk, while falling-to-flat rotational curves are indicative of a massive galactic bulge (e.g. NGC224, NGC 2841, see Kent 1986 and Kent 1987).

## 5. ROTATIONAL DRAG FORCE EFFECT ON GALACTIC MORPHOLOGY

### 5.1. Off-Plane Rotational Drag Force

Rotational drag force on galactic level is caused by the rotation of the galactic center of mass, to which the major contributors are the co-rotating galactic bulge and possibly the super-massive black hole located in the bulge's center (presuming that the black holes produce rotational drag force, or that the super-massive objects at galactic cores while dense and compact are not really black holes). It is believed that virtually every galaxy harbors a super-massive black hole in its center; see for example Magorrian et al. (1998) and Kormendy & Richstone (1995). However, further exploration is necessary to determine the dominant cause of the rotational drag force on the galactic level.

While typically the rotation plane of the galactic center of mass will coincide with the rotational plane of the galactic disk, it is reasonable to expect that some events in galactic history, such as catastrophic collisions in the bulge or mergers will misalign the rotational axis of the galactic center of mass from the rotational axis of the galactic disk. In this event the rotational drag force will not longer act in galactic plane but would rather receive an off-plane component causing massive galactic warp or even 'star peeling' off the galactic rim. Note that the star peeling will occur in opposite directions on the both sides of the galactic disk with one peeled-off spiral arm bending above the disk and another spiral arm bending below the disk. When viewed from the side the resulting three-dimensional structure will look like a barred galaxy although the bar in reality is a disk observed at an angle.

Modeling of the galactic warps and galactic bar structure on the basis of the rotational drag force theory is discussed in depth in section 6 of this paper.

### 5.2. Rotational Drag Force at Non-Equilibrium State

The expressions (10) and (14) for rotational drag force acceleration and rotational velocity model are derived for an *equilibrium* state. The equilibrium state implies that:

-   the only existing rotational drag force component is radial and no tangential component exists;

-   orbital velocities are stationary and do not change with time.

While the equilibrium state can be expected for older galaxies at late stages of evolution, young galaxies will exhibit quite different behavior. Indeed, on early stages of galactic evolution while the co-rotation in galactic bulge is still developing and / or the central black hole has not yet been formed, the galaxy is not expected to produce a rotational drag force, as there is no coherent rotational pattern in the galactic center of mass. Chaotic bulge stellar orbits and Newtonian rotational velocity curve describe such galaxy.

However, as orbits in the bulge are getting organized towards co-rotation and the central black hole grows more massive, rotational drag force is expected to arise forcing stars in plane of the center of mass rotation to accelerate and move onto higher orbits resulting in increase of their orbital velocity and development of galactic disk. An exchange of angular momentum will occur between the galactic bulge / central black hole and the galactic disk. With time an equilibrium state will develop when majority of stars in the galactic disk attain equilibrium orbital velocities described by equation (14). At that point tangential component of the rotational drag force will cease to exist. In other words rotational drag force exhibits viscous behavior.

To derive an expression for rotational drag force in non-equilibrium state let's consider a star orbiting the galactic center of mass with momentary velocity vector **v** and its orbit tilted from the rotational drag force plane at an angle *x*. Then the expression for the rotational drag force acceleration exerted on the star is:

$$|a_{rot}| = \frac{(v'-wr')^2}{r'(1+g^2(v'-wr')^2)} \tag{22}$$

where $r'$ – projection of the star's radius-vector $\mathbf{r}$ on the plane of rotation of the galactic center of mass, $v'$ – projection of the star's velocity vector $\mathbf{v}$ on the tangent-vector $\mathbf{T}$ to a circular orbit residing in the plane of rotation of the galactic center of mass:

$$r' = |\mathbf{r}| |\sin x| \tag{23}$$

$$v' = \mathbf{v} \, \mathbf{T} \tag{24}$$

To derive radial and tangential projections of the drag force-induced acceleration $a_R$ and $a_T$ consider the following reasoning. When the rotation in the galactic center of mass is chaotic and no rotational drag force is produced stellar orbits in the galactic disk will be described by purely Newtonian law of gravity. As soon as the rotation in the galactic bulge becomes coherent the rotational drag force kicks in and starts accelerating stars in the galactic disk until they attain equilibrium velocities described by equation (14). In the case of flat rotational curve the equilibrium velocity for stars in the disk is $g^{-1}$, while the stars in the bulge co-rotate and do not experience the drag force. Thus when $|\mathbf{v}| < g^{-1}$ both tangential and radial projections of the drag force are present with tangential projection ceasing to exist when $|\mathbf{v}| \to g^{-1}$. On the other hand when a star is stationary (i.e. $|\mathbf{v}| = 0$) the radial projection would not exist with only tangential drag force accelerating the star.

The following equations summarize the viscous behavior of the rotational drag force:

$$a_R \to \begin{cases} 0, |\mathbf{v}| \mapsto 0; \\ |a_{rot}|, |\mathbf{v}| \mapsto g^{-1}. \end{cases} \tag{25}$$

and

$$a_T \to \begin{cases} |a_{rot}|, |\mathbf{v}| \mapsto 0; \\ 0, |\mathbf{v}| \mapsto g^{-1}. \end{cases} \tag{26}$$

Which is equivalent to

$$a_R = |a_{rot}| \sin q \tag{27}$$

and

$$a_T = |a_{rot}| \cos q \tag{28}$$

where

$$\sin q = \frac{2\sqrt{g^{-1}v'}}{g^{-1}+v'} \tag{29}$$

and

$$\cos q = \frac{g^{-1}-v}{g^{-1}+v'} \tag{30}$$

The expressions (29) and (30) satisfy the conditions (25) and (26) and the condition that $\sin^2 q + \cos^2 q = 1$.

The derived expressions are sufficient for conducting numerical modeling of the effects of rotational drag force on galactic morphology in the non-equilibrium state. The non-equilibrium state includes a scenario when the axis of rotation of the galactic center of mass does not coincide with axis of rotation of the galactic disk. The latter situation is equivalent to the non-equilibrium state when the axes do coincide but the stellar velocities in disk are purely Newtonian because the projections of the rotational velocities of stars in the galactic disk on the plane of rotation of the galactic center of mass will differ from the equilibrium velocity $g^{-1}$.

## 6. STELLAR DYNAMICS SIMULATION WITH ROTATIONAL DRAG FORCE

### 6.1. Stellar Dynamics Simulation Software

To model the effect of rotational drag force on galactic morphology I have developed specialized software for stellar dynamics simulation and real-time 3D visualization of galactic morphology (C++, MS Windows, DirectX). The source code and the executable module are available upon request.

The software represents a galaxy as a bulge + disk system of roughly 10,000 stars each. For improved performance the software computational algorithm does not account for individual gravitational interaction between stars, which is considered weak to be of immediate interest, but rather calculates stellar motion using Newtonian gravitational potentials for bulge and disk with rotational drag force modeled as additional acceleration acting on stars.

Each simulation session begins with an initial galactic configuration corresponding to a disk galaxy of type S0-fig. 4-with stellar velocities calculated using gravitational Newtonian potential. Then an amplified rotational drag force forming a specified angle with the galactic disk is introduced and its effect on the stellar dynamics and galactic morphology is visualized.

Drag force amplification is necessary to reduce the simulation time and to avoid rotational velocity error accumulation and shot noise in the galactic bulge where the gravitational potential is strong. Also, as the galactic shape deviates from the original the simulation becomes less accurate as the original Newtonian gravitational potential is assumed throughout the entire session. Overall the developed model provides sufficient means for studying of the immediate impact of the rotational drag force on galactic morphology and reveals qualitative changes in galactic disk structure.

### 6.2. Galactic Model

For galactic bulge I employ an oblate spheroid model with mass-density described by Gaussian-like distribution, derived from Dwek et al. 1995:

$$\boldsymbol{r}_{bulge}(R) = \frac{\boldsymbol{r}_0}{R^2}\exp[-\frac{1}{2}R^2]\qquad(31)$$

$$R^2 = \frac{x^2+y^2}{r_0^2}+\frac{z^2}{z_0^2}\equiv\frac{r^2}{r_0^2}+\frac{z^2}{z_0^2}\qquad(32)$$

Where $r_0 = 0.91$ kpc and $z_0 = 0.51$ kpc.

For galactic disk I employ an exponential disk model found in Wainscoat 1992:

$$\boldsymbol{r}_{disk}(r,z) = \boldsymbol{r}_1\exp[\frac{(R_0-r)}{h}-\frac{|z|}{h_z}]\qquad(33)$$

Where $R_0 = 8.5$ kpc and $h_z = 0.2$ kpc.

The values for $\boldsymbol{r}_0$ and $\boldsymbol{r}_1$ are selected to yield equivalent masses for the bulge and for the disk, while the total mass of the galaxy is set to be 100 billion solar masses.

With the galactic bulge radius of 5 kpc and the galactic disk radius and thickness of 15 kpc and 1 kpc respectively the resulting galactic model is somewhat similar to Milky Way without the molecular ring and with no spiral arms – fig.4.

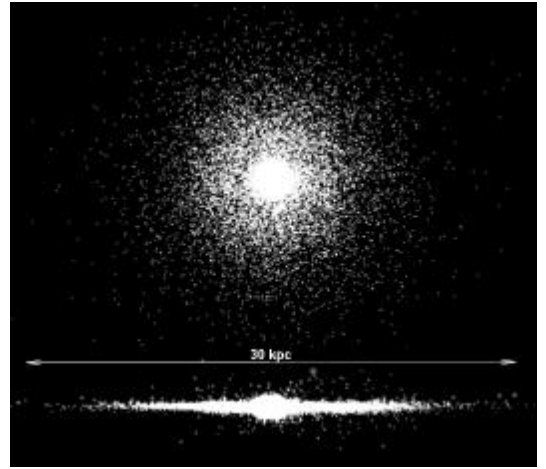

Fig. 4.-Starting disk galaxy shape for rotational drag force modeling, galactic disk diameter is 30 kpc.

### 6.3. Origination of Spiral Structure

The first experiment for modeling the effect of the rotational drag force on the galactic disk structure was conducted with the rotational drag force acting strictly in plane of the galactic disk. When subjected to the rotational drag force disk stars began accelerating from the initial Newtonian velocities towards the equilibrium velocity, which eventually lead the galaxy to developing spiral arm structure depicted on fig. 5, a totally unexpected result. The spiral arm structure remained stable and exhibited slow winding around the galactic core. The observed spiral arm winding occurred.

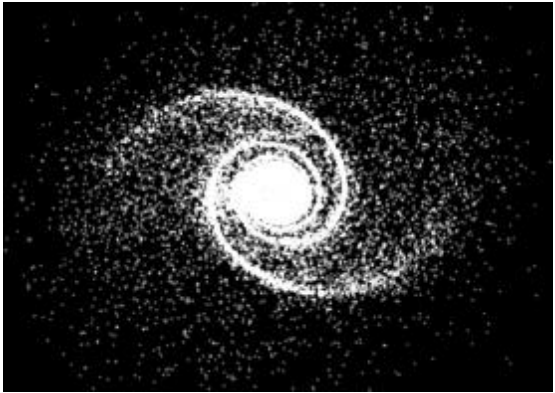

Fig. 5.-A disk galaxy developing spiral structure with rotational drag force is acting in plane of the galactic disk.

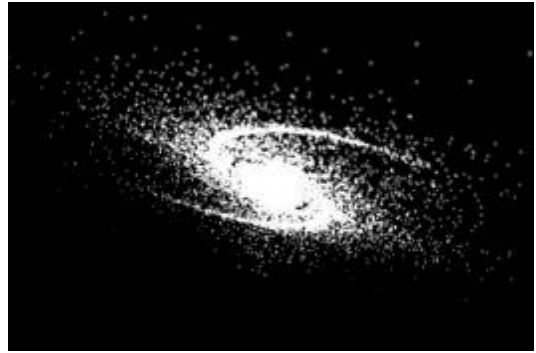

Fig. 6.-Early stage of a disk galaxy developing spiral structure (seen from the side). The rotational drag force forms an extreme 90° angle with galactic disk and peels stars off the galactic rim producing two spiral arms to bent above and below the plane of the galactic disk.

### 6.4. Galactic Bars

Another experiment was conducted with the rotational drag force forming an extreme 90° angle with the galactic plane. Spiral arms were once again produced although through the effect of 'star peeling' off the galactic rim. Because of the extreme angle of the rotational drag force the produced spiral arms were bent above and below the galactic plane and formed a three-dimensional spiral rather than two-dimensional spiral as in the case of in-pane rotational drag force-fig. 6.

When viewed from the side the resulting 3D galactic morphology strikingly reminds galactic bar structure, which further evolves into an extremely warped spiral as shown of fig. 7. The latter features abnormally large spiral arms reaching far above and below the initial galactic plane. Curiously, but the obtained image looks remarkably similar to the M33 spiral galaxy, which means that if the rotational drag force theory is correct then the spiral arms of M33 may be not at all equidistant from Milky Way as they form a grossly distorted three-dimensional structure, which only *appears* to be two-dimensional because it seen from the pole.

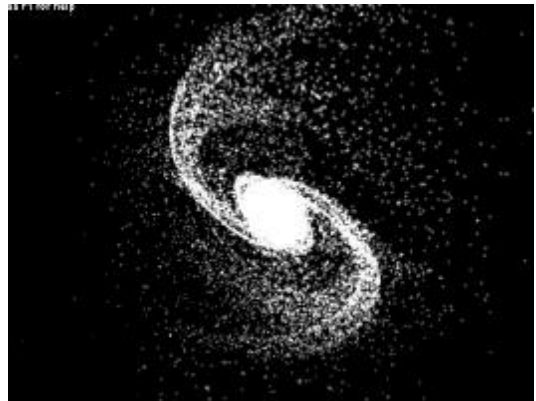

(a)

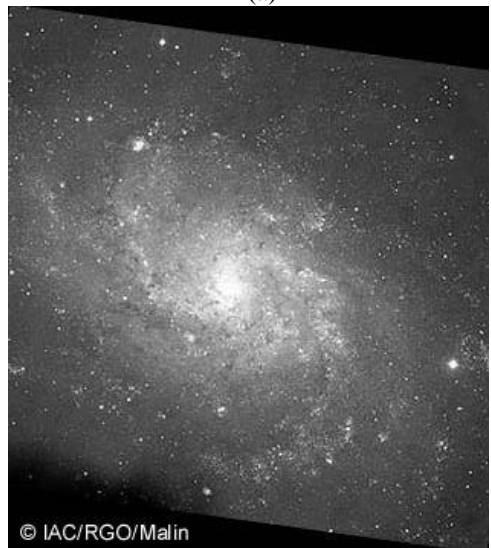

(b)

Fig. 7.-What looks like a barred galaxy (a) is in fact a grotesquely warped spiral. The galaxy was subjected to strong rotational drag force forming a 90° angle with the galactic disk. Extended simulation time allowed the warp to grow into to an exaggerated proportion. The outcome looks remarkably similar to M33 (b). M33 image is a courtesy of David Malin.

### 6.5. Galactic Warps

Yet another series of experiments was conducted with the moderate rotational drag force forming various angles between 15° and 90° with the galactic plane. Early stages of galactic evolution produced galactic warps-fig. 8-with higher rotational drag force angles producing stronger warps. The warps remained stable when stellar velocities approached equilibrium value.

The observed behavior suggests that rotational drag force might be a better explanation for galactic warps than extragalactic magnetic fields (Battaner & Jiménez-Vicente 1990), which are considered to be too strong for intergalactic medium (Binney 1991 and Combes 1994), or asymmetric halo model (Binney 1992), which suffers from an initial-value problem.

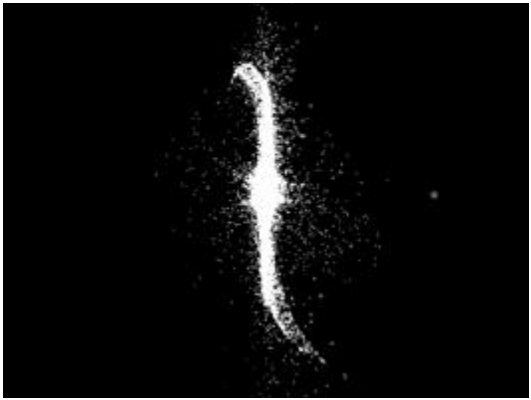

Fig. 8.- Edge-on view of a galactic warp produced by moderate off-plane rotational drag force on early stage of the galactic evolution.

### 7. CONCLUSION

Flat rotational velocity curves of galaxies and numerous features of galactic morphology such as spiral arms, galactic bars and galactic warps suggest that an unknown viscous force is acting on stars on galactic level. In this paper I have hypothesized that such viscous force is a rotational drag force originating in galactic bulge. Unlike Modified Newtonian Dynamics the rotational drag force is fully compatible with Newtonian dynamics and is based on the overlooked effect of quantum particle tunneling through gravitational energy fluctuations in the rotating gravitational field.

The proposed rotational drag force appears to be a successful tool for explaining flat rotational curves of galaxies without the need for dark matter. Preliminary stellar dynamics simulation results provide a good indication that the observed galactic morphology phenomena such as spiral arms, bars and warps are much easier to understand with the help of the rotational drag force.

Clearly, detailed quantitative analysis of the rotational drag force implications as well as more accurate theoretical and computer simulation of its effects on stellar dynamics and galactic morphology is in order. I call for such analysis to be conducted to help validate and further develop the rotational drag force theory.

I thank Joe Hershberger for assistance with simulation software development.